\documentclass[a4paper,11pt]{article}
\usepackage[T1]{fontenc} % if needed
\usepackage{float} % if needed

\makeatletter
\gdef\@fpheader{ }
\gdef\@journal{ }
\makeatother
\RequirePackage{amsmath}
\RequirePackage{amssymb}
\RequirePackage{epsfig}
\RequirePackage{graphicx}
\RequirePackage[numbers,sort&compress]{natbib}
\RequirePackage{color}
\RequirePackage[colorlinks=true
,urlcolor=blue
,anchorcolor=blue
,citecolor=blue
,filecolor=blue
,linkcolor=blue
,menucolor=blue
,pagecolor=blue
,linktocpage=true
,pdfproducer=medialab
,pdfa=true
]{hyperref}

\newif\ifnotoc\notocfalse
\newif\ifemailadd\emailaddfalse
\newif\iftoccontinuous\toccontinuousfalse
\makeatletter
\def\@subheader{\@empty}
\def\@keywords{\@empty}
\def\@abstract{\@empty}
\def\@xtum{\@empty}
\def\@dedicated{\@empty}
\def\@arxivnumber{\@empty}
\def\@collaboration{\@empty}
\def\@collaborationImg{\@empty}
\def\@proceeding{\@empty}
\def\@preprint{\@empty}

\newcommand{\subheader}[1]{\gdef\@subheader{#1}}
\newcommand{\keywords}[1]{\if!\@keywords!\gdef\@keywords{#1}\else%
\PackageWarningNoLine{\jname}{Keywords already defined.\MessageBreak Ignoring last definition.}\fi}
\renewcommand{\abstract}[1]{\gdef\@abstract{#1}}
\newcommand{\dedicated}[1]{\gdef\@dedicated{#1}}
\newcommand{\arxivnumber}[1]{\gdef\@arxivnumber{#1}}
\newcommand{\proceeding}[1]{\gdef\@proceeding{#1}}
\newcommand{\xtumfont}[1]{\textsc{#1}}
\newcommand{\correctionref}[3]{\gdef\@xtum{\xtumfont{#1} \href{#2}{#3}}}
\newcommand\jname{JHEP}

\newcommand\preprint[1]{\gdef\@preprint{\hfill #1}}

\makeatother

\newcommand\note[2][]{%
\if!#1!%
\stepcounter{footnote}\footnotetext{#2}%
\else%
{\renewcommand\thefootnote{#1}%
\footnotetext{#2}}%
\fi}

\makeatletter
\newtoks\auth@toks
\renewcommand{\author}[2][]{%
 \if!#1!%
  \auth@toks=\expandafter{\the\auth@toks#2\ }%
 \else
  \auth@toks=\expandafter{\the\auth@toks#2$^{#1}$\ }%
 \fi
}
\makeatother
\makeatletter
\newtoks\affil@toks\newif\ifaffil\affilfalse
\newcommand{\affiliation}[2][]{%
\affiltrue
 \if!#1!%
  \affil@toks=\expandafter{\the\affil@toks{\item[]#2}}%
 \else
  \affil@toks=\expandafter{\the\affil@toks{\item[$^{#1}$]#2}}%
 \fi
}
\makeatother
\makeatletter
\newtoks\email@toks\newcounter{email@counter}%
\newcommand{\emailAdd}[1]{%
\emailaddtrue%
\ifnum\theemail@counter>0\email@toks=\expandafter{\the\email@toks, \@email{#1}}%
\else\email@toks=\expandafter{\the\email@toks\@email{#1}}%
\fi\stepcounter{email@counter}}
\newcommand{\@email}[1]{\href{mailto:#1}{\tt #1}}
\makeatother

\makeatletter
\newcommand*\collaboration[1]{\gdef\@collaboration{#1}}
\newcommand*\collaborationImg[2][]{\gdef\@collaborationImg{#2}}
\makeatletter
\newcommand\afterLogoSpace{\smallskip}
\newcommand\afterSubheaderSpace{\vskip3pt plus 2pt minus 1pt}
\newcommand\afterProceedingsSpace{\vskip21pt plus0.4fil minus15pt}
\newcommand\afterTitleSpace{\vskip23pt plus0.06fil minus13pt}
\newcommand\afterRuleSpace{\vskip23pt plus0.06fil minus13pt}
\newcommand\afterCollaborationSpace{\vskip3pt plus 2pt minus 1pt}
\newcommand\afterCollaborationImgSpace{\vskip3pt plus 2pt minus 1pt}
\newcommand\afterAuthorSpace{\vskip5pt plus4pt minus4pt}
\newcommand\afterAffiliationSpace{\vskip3pt plus3pt}
\newcommand\afterEmailSpace{\vskip16pt plus9pt minus10pt\filbreak}
\newcommand\afterXtumSpace{\par\bigskip}
\newcommand\afterAbstractSpace{\vskip16pt plus9pt minus13pt}
\newcommand\afterKeywordsSpace{\vskip16pt plus9pt minus13pt}
\newcommand\afterArxivSpace{\vskip3pt plus0.01fil minus10pt}
\newcommand\afterDedicatedSpace{\vskip0pt plus0.01fil}
\newcommand\afterTocSpace{\bigskip\medskip}
\newcommand\afterTocRuleSpace{\bigskip\bigskip}
\newlength{\affiliationsSep}
\newcommand\beforetochook{\pagestyle{myplain}\pagenumbering{roman}}

\DeclareFixedFont\trfont{OT1}{phv}{b}{sc}{11}

\renewcommand\maketitle{
\pagestyle{empty}
\thispagestyle{titlepage}
\setcounter{page}{0}
\noindent{\small\scshape\@fpheader}\@preprint\par

\afterLogoSpace
\if!\@subheader!\else\noindent{\trfont{\@subheader}}\fi
\afterSubheaderSpace
\if!\@proceeding!\else\noindent{\sc\@proceeding}\fi
\afterProceedingsSpace
{\LARGE\flushleft\sffamily\bfseries\@title\par}
\afterTitleSpace
\hrule height 1.5\p@%
\afterRuleSpace
\if!\@collaboration!\else
{\Large\bfseries\sffamily\raggedright\@collaboration}\par
\afterCollaborationSpace
\fi
\if!\@collaborationImg!\else
{\normalsize\bfseries\sffamily\raggedright\@collaborationImg}\par
\afterCollaborationImgSpace
\fi
{\bfseries\raggedright\sffamily\the\auth@toks\par}
\afterAuthorSpace
\ifaffil\begin{list}{}{%
\setlength{\leftmargin}{0.28cm}%
\setlength{\labelsep}{0pt}%
\setlength{\itemsep}{\affiliationsSep}%
\setlength{\topsep}{-\parskip}}
\itshape\small%
\the\affil@toks
\end{list}\fi
\afterAffiliationSpace
\ifemailadd %% if emailadd is true
\noindent\hspace{0.28cm}\begin{minipage}[l]{.9\textwidth}
\begin{flushleft}
\textit{E-mail:} \the\email@toks
\end{flushleft}
\end{minipage}
\else %% if emailaddfalse do nothing
\PackageWarningNoLine{\jname}{E-mails are missing.\MessageBreak Plese use \protect\emailAdd\space macro to provide e-mails.}
\fi
\afterEmailSpace
\if!\@xtum!\else\noindent{\@xtum}\afterXtumSpace\fi
\if!\@abstract!\else\noindent{\renewcommand\baselinestretch{.9}\textsc{Abstract:}}\ \@abstract\afterAbstractSpace\fi
\if!\@keywords!\else\noindent{\textsc{Keywords:}} \@keywords\afterKeywordsSpace\fi
\if!\@arxivnumber!\else\noindent{\textsc{ArXiv ePrint:}} \href{http://arxiv.org/abs/\@arxivnumber}{\@arxivnumber}\afterArxivSpace\fi
\if!\@dedicated!\else\vbox{\small\it\raggedleft\@dedicated}\afterDedicatedSpace\fi
\ifnotoc\else
\iftoccontinuous\else\newpage\fi
\beforetochook\hrule
\tableofcontents
\afterTocSpace
\hrule
\afterTocRuleSpace
\fi
\setcounter{footnote}{0}
\pagestyle{myplain}\pagenumbering{arabic}
} % close the \renewcommand\maketitle{

\renewcommand{\baselinestretch}{1.1}\normalsize
\divide\@tempdima\baselineskip
\@tempcnta=\@tempdima
\skip\@mpfootins = \skip\footins
\renewcommand{\@dotsep}{10000}

\newcommand\ps@myplain{
\pagenumbering{arabic}
\renewcommand\@oddfoot{\hfill-- \thepage\ --\hfill}
\renewcommand\@oddhead{}}
\let\ps@plain=\ps@myplain

\newcommand\ps@titlepage{\renewcommand\@oddfoot{}\renewcommand\@oddhead{}}

\numberwithin{equation}{section}

\renewcommand\section{\@startsection{section}{1}{\z@}%
                  {-3.5ex \@plus -1.3ex \@minus -.7ex}%
                  {2.3ex \@plus.4ex \@minus .4ex}%
                  {\normalfont\large\bfseries}}
\renewcommand\subsection{\@startsection{subsection}{2}{\z@}%
                  {-2.3ex\@plus -1ex \@minus -.5ex}%
                  {1.2ex \@plus .3ex \@minus .3ex}%
                  {\normalfont\normalsize\bfseries}}
\renewcommand\subsubsection{\@startsection{subsubsection}{3}{\z@}%
                  {-2.3ex\@plus -1ex \@minus -.5ex}%
                  {1ex \@plus .2ex \@minus .2ex}%
                  {\normalfont\normalsize\bfseries}}
\renewcommand\paragraph{\@startsection{paragraph}{4}{\z@}%
                  {1.75ex \@plus1ex \@minus.2ex}%
                  {-1em}%
                  {\normalfont\normalsize\bfseries}}
\renewcommand\subparagraph{\@startsection{subparagraph}{5}{\parindent}%
                  {1.75ex \@plus1ex \@minus .2ex}%
                  {-1em}%
                  {\normalfont\normalsize\bfseries}}

\def\fnum@figure{\textbf{\figurename\nobreakspace\thefigure}}
\def\fnum@table{\textbf{\tablename\nobreakspace\thetable}}

\long\def\@makecaption#1#2{%
 \vskip\abovecaptionskip
 \sbox\@tempboxa{\small #1. #2}%
 \ifdim \wd\@tempboxa >\hsize
  \small #1. #2\par
 \else
  \global \@minipagefalse
  \hb@xt@\hsize{\hfil\box\@tempboxa\hfil}%
 \fi
 \vskip\belowcaptionskip}

\renewenvironment{thebibliography}[1]{%
\begin{oldthebibliography}{#1}%
\small%
\raggedright%
\setlength{\itemsep}{5pt plus 0.2ex minus 0.05ex}%
}%
{%
\end{oldthebibliography}%
}
\begin{document}

%%%%%%%%%%%%%%%%%%炎籾匈%%%%%%%%%%%%%%%%%%%%%%%%%%%%%&&&&&&&&&&&&&&&&&&&&&&

%\title{\boldmath How is the Information of the Return Contained in the Financial Time Series}
\title{\boldmath Why Existing Machine Learning Methods Fails At
Extracting the Information of Future Returns Out of Historical Sctock Prices
: the Curve-Shape-Feature and Non-Curve-Shape-Feature Modes}
% more complex case: 4 authors, 3 institutions, 2 footnotes

\author[a]{Jia-Yao Yang,}
\author[a,1]{Hao Zhu,}\note{Hao Zhu and Jia-Yao Yang contributed equivalently to this work.}
\author[a,2]{Yue-Jie Hou,}
\author[b,2]{Ping Zhang,}\note{zhangping@cueb.edu.cn. Corresponding author}
\author[a,3]{and Chi-Chun Zhou}\note{zhouchichun@dali.edu.cn. Corresponding author}

% The "\note" macro give a warning: "Ignoring empty anchor..."
% you can safely ignore it.

\affiliation[a]{School of Engineering, Dali University, Dali, Yunnan 671003, PR China}
\affiliation[b]{School of Finance, Capital University of Economics and Business, Beijing 100070, P. R. China}

%\affiliation[c]{DP School}

% e-mail addresses: one for each author, in the same order as the authors
%\emailAdd{Ccc@one.edu.cn}
%\emailAdd{second@asas.edu}
%\emailAdd{daiwusheng@tju.edu.cn}
%\emailAdd{fourth@one.univ}

%\title{\boldmath A title with some math: $x=1$}
%% %simple case: 2 authors, same institution
%% \author{A. Uthor}
%% \author{and A. Nother Author}
%% \affiliation{Institution,\\Address, Country}

% more complex case: 4 authors, 3 institutions, 2 footnotes
%\author[a,b,1]{F. Irst,\note{Corresponding author.}}
%\author[c]{S. Econd,}
%\author[a,2]{T. Hird\note{Also at Some University.}}
%\author[a,2]{and Fourth}

% The "\note" macro give a warning: "Ignoring empty anchor..."
% you can safely ignore it.

%\affiliation[a]{One University,\\some-street, Country}
%\affiliation[b]{Another University,\\different-address, Country}
%\affiliation[c]{A School for Advanced Studies,\\some-location, Country}

% e-mail addresses: one for each author, in the same order as the authors
%\emailAdd{first@one.univ}
%\emailAdd{second@asas.edu}
%\emailAdd{third@one.univ}
%\emailAdd{fourth@one.univ}
%\date{date}

\abstract{The financial time series analysis is important access 
to touch the complex laws of financial markets.
Among many goals of the financial time series analysis,
one is to construct a model 
that can extract the information of the future return out 
of the known historical stock data, such as stock price, financial news, and e.t.c.
To design such a model, prior knowledge on how the future return 
is correlated with the historical stock prices is needed.
In this work, we focus on the issue: in what mode the future return 
is correlated with the historical stock prices.
We manually design several financial time series where the
future return is correlated with the historical stock prices in pre-designed
modes, namely the curve-shape-feature (CSF) and 
the non-curve-shape-feature (NCSF) modes. In the CSF mode, 
the future return can be extracted from the curve shapes of the historical
stock prices.
By applying various kinds of existing 
algorithms on those pre-designed time series and 
real financial time series, we show that:
(1) the major information of the future return is not contained in 
the curve-shape features of historical stock prices.
That is, the future return is not mainly correlated with the historical stock prices
in the CSF mode.
(2) Various kinds of existing machine learning algorithms
are good at extracting the curve-shape features in the historical stock prices 
and thus are inappropriate for financial time series analysis 
although they are successful in the image recognition and natural language processing.
That is, existing machine learning methods that are good at handling the CSF series will fail at 
extracting the information of future returns out of historical stock prices.
New models handling the NCSF series are needed in 
the financial time series analysis.}
%\keywords{}
\keywords{Machine learning, Classification, Feature extraction, Financial time series. Deep learning}
\maketitle
\flushbottom
%%%%%%%%%%%%%%%%%%炎籾匈潤崩%%%%%%%%%%%%%%%%%%%%%%%%%%%%%&&&&&&&&&&&&&&&&&&&

%%%%%%%%%%屎猟蝕兵

\section{Introduction} 
The financial market is indeed a complex and giant system 
\cite{tsay2005analysis,mills2008econometric,andersen2009handbook}.
The financial time series analysis is important access 
to touch the complex laws of financial markets 
\cite{tsay2005analysis,mills2008econometric,andersen2009handbook}. 
Among many goals of the financial time series analysis, 
one is to construct a model 
that can extract the information of the future return out 
of the already known historical data, such as stock prices, financial news, 
economic events, and political events
\cite{tay2001application,krollner2010financial,al2013predicting,adhikari2014combination,sezer2020financial}. 
 
Before constructing a model that can extract the information of the future return out 
of the historical data, researchers need the prior-knowledge of the markets
and investigate issues such as whether the 
information of the future return contained in the already known historical data? Or,
is there a correlation between the future return
and the historical stock prices?
According to the effective market hypothesis 
(EMH) \cite{basu1977investment,malkiel1989efficient,malkiel2003efficient}, 
stocks always trade at their fair value on exchanges, thus, no investors
can outperform the overall market through stock selection or market timing, 
the higher returns can only be obtained by purchasing riskier investments.
Therefore, according to EMH, there is no correlation between the current price 
and the future price of the stock market, I.e., 
any change in the stock price is completely independent of the past price.
However, there are different opinions on the issue of EMH 
\cite{basu1977investment,malkiel2003efficient,degutis2014efficient,hamid2017testing} and 
higher returns can be obtained by technical analysis such as 
expert stock selection and market timing in real investments 
\cite{blume1994market,murphy1999technical,wong2003rewarding,park2007we,kirkpatrick2010technical}.
In this regard, we cannot deny that the historical data of financial 
markets contains information of the future return.
Therefore, it is of great significance to analyze the hidden features and 
laws of historical financial data, including stock prices, financial news, 
economic events, and political events. 

Stock prices are typical data that are accessible and intuitive.
There are research studying how to extract future return from 
the known historical stock prices.
Before the popularity of machine learning methods, 
researchers investigate the financial time series
from the perspective of analytical and statistical methods. For example, 
the fractional market shows that the financial market has the characteristics of fractional and 
non-linearity \cite{scheinkman1989nonlinear, peters1994fractal,peters1996chaos,laskin2000fractional}. 
Spectrum analysis methods \cite{rostan2018versatility}, 
such as the Fourier transform \cite{kawagoe2002similarity,cherubini2010fourier} 
and the wavelet transform \cite{kawagoe2002similarity,zhang2006unsupervised, addison2017illustrated} 
are applied to the financial time series analysis.
The auto regressive model (AR), the moving average model (MA), 
the auto regressive and moving average model (AR-MA),
and e.t.c., are proposed to modeling the market \cite{tsay2005analysis,mills2008econometric,andersen2009handbook}.
Besides is directly forecasting the future return, 
the generalized autoregressive conditional heteroskedasticity model (GARCH) 
is proposed to model and forecast 
conditional mean and volatility \cite{bauwens2006multivariate}. 
Other hybrid models such as ARMA-GARCH \cite{ling2003asymptotic,francq2004maximum} and 
its improved ARMA-GARCH-M \cite{liu2011comprehensive} model are proposed . 
However, the statistical method is not suitable for actively discovering various potential 
rules from numerous data \cite{hand2007principles} 

With the popularity of machine learning methods, researchers
use the machine learning methods to investigate the financial time series. 
For example, the support vector machines (SVM) \cite{tay2001application,kim2003financial}
, the recurrent neural network (RNN) \cite{madan2018predicting} 
and it's generalization long-short term memory (LSTM) 
\cite{wu2018adaboost,cao2019financial} network, and the convolutional neural network (CNN)
\cite{livieris2020cnn,mehtab2020analysis}
are applied in financial time series forecasting.
A multi-scale recurrent convolutional neural network (MSTD-RCNN) \cite{2019Multi} 
is proposed and proved to improve the accuracy of data prediction.
The hybrid models which is the combination of the statistical method and the machine learning method
such as multi forecast model of ARIMA and artificial neural 
network (ANN) \cite{2014A}, the FEPA model (FTS-EMD-PCA-ANN) \cite{Zhang2016A},
nonlinear autoregressive neural network model \cite{2020Nonlinear} are used in modeling 
the financial time series. 
Beyond the model structure designing, 
other researchers focus on the data. For example,
Ref. \cite{2020Evaluating} evaluates several augmentation methods applied to stocks datasets using two 
state-of-the-art deep learning models and show that several augmentation
methods significantly improve financial performance 
when used in combination with a trading strategy.

The majority of the research focused on how to improve the model's behavior
in forecasting the future return and most of the proposed models report a improving 
in the forecasting accuracy.
However, there are research report a failure of the deep learning approach
on financial time series analysis \cite{dingli2017financial}.
In our opinions, the prior-knowledge on how 
the future stock price or return is correlated with the historical stock price 
should be obtained before designing an effect algorithm or model. 
Unfortunately,
to our knowledge, not many research concern
the question about in what mode the future return is correlated with the historical stock price.
Or, how is the information of the 
future return contained in the already known stock price.
With the absence of the prior-knowledge, one might become 
blind in choosing models.

In this paper, under the assumption that the market is not completely effective,
we focus on the issue: in what mode the future return 
is correlated with the historical stock prices.
We manually design several financial time series where the
future return is correlated with the historical stock prices in pre-designed
modes, namely the curve-shape-feature (CSF) and 
the non-curve-shape-feature (NCSF) modes. In the CSF mode, 
the future return can be extracted from the curve shapes of the historical
stock prices. In the NCSF mode, the information of future return is not contained 
in the curve shape of historical stock prices. 
By applying various kinds of existing 
algorithms on those pre-designed time series, 
we find that various kinds of 
existing models only perform well on 
the CSF mode series and fail on the NCSF mode series.
By comparing the behavior of the same algorithm on the CSF mode series, NCSF mode series, and
real financial time series, 
we conclude that:
(1) the major information of the future return is not contained in 
the curve-shape features of historical stock prices.
That is, the future return is not mainly correlated with the historical stock prices
in the CSF mode.
(2) Various kinds of existing machine learning algorithms
are good at extracting the curve-shape features in the historical stock prices 
and thus are inappropriate for financial time series analysis 
although they are successful in the image recognition, nature language processing, and e.t.c.
It points out that beyond the existing models, new models that can extract 
non-curve-shape features are needed in 
the financial time series analysis.

This paper is organized as follows:
In Sec. 2, the pre-designed time series, including 
the CSF mode series and NCSF mode series, are introduced.
In Sec. 3, we firstly give a brief review on various existing algorithms 
and secondly apply them on the CSF mode series, NCSF mode series, and the real series.
In Sec. 4, we analyze the results.
Conclusions and discussions are given in Sec. 5. 

\section{Pre-designed series: the CSF mode series and NCSF mode series}
The prior-knowledge that in what mode 
the future return is correlated with the historical stock prices 
is important in financial time series analysis. 
In this section, we manually design several financial 
time series where the future return is correlated 
with the historical stock prices in pre-designed
modes, namely the curve-shape-feature (CSF) and 
the non-curve-shape-feature (NCSF) modes. 

\subsection{The CSF mode series}
In this section, we introduce the CSF mode series. In the CSF mode series 
the future return can be extracted from the curve shapes of the historical
stock price.

\begin{figure}[ht]
\centering
\includegraphics[width=1.0\textwidth]{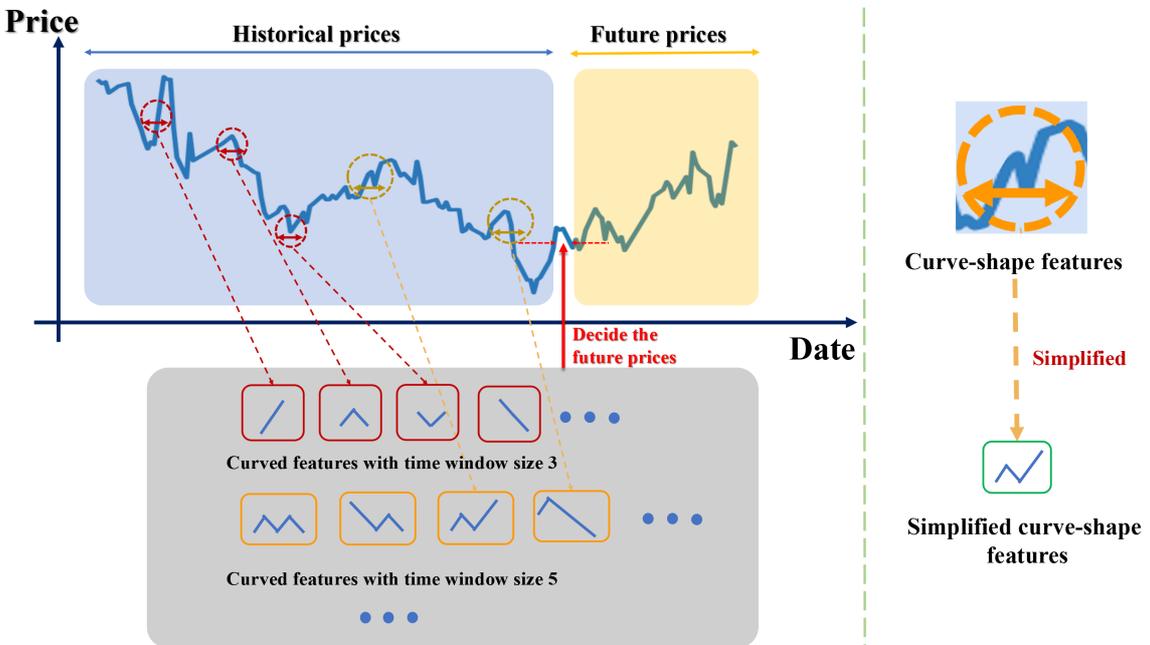}
\caption{Examples of the curve-shape features and simplified curve-shape features.
In the simplified curve-shape features only the trend of increase and decrease
is considered.}
\label{curved_feature}
\end{figure}

\textit{The curve-shape features and simplified curve-shape features}. 
In the historical stock price, 
the curve-shape features are morphological shapes occurring
in a window with given size. 
There is a large amount of the curve-shape features. In this work,
we simplify the curve-shape features by ignoring the magnitude of the 
stock prices and only considering the trend of increase and decrease, 
as shown in Fig. \ref{curved_feature}. 
In the following discussion, we consider the simplified curve-shape features only
and make no distinguish between the curve-shape 
features and simplified curve-shape features.

\begin{figure}[H]
\centering
\includegraphics[width=1.0\textwidth]{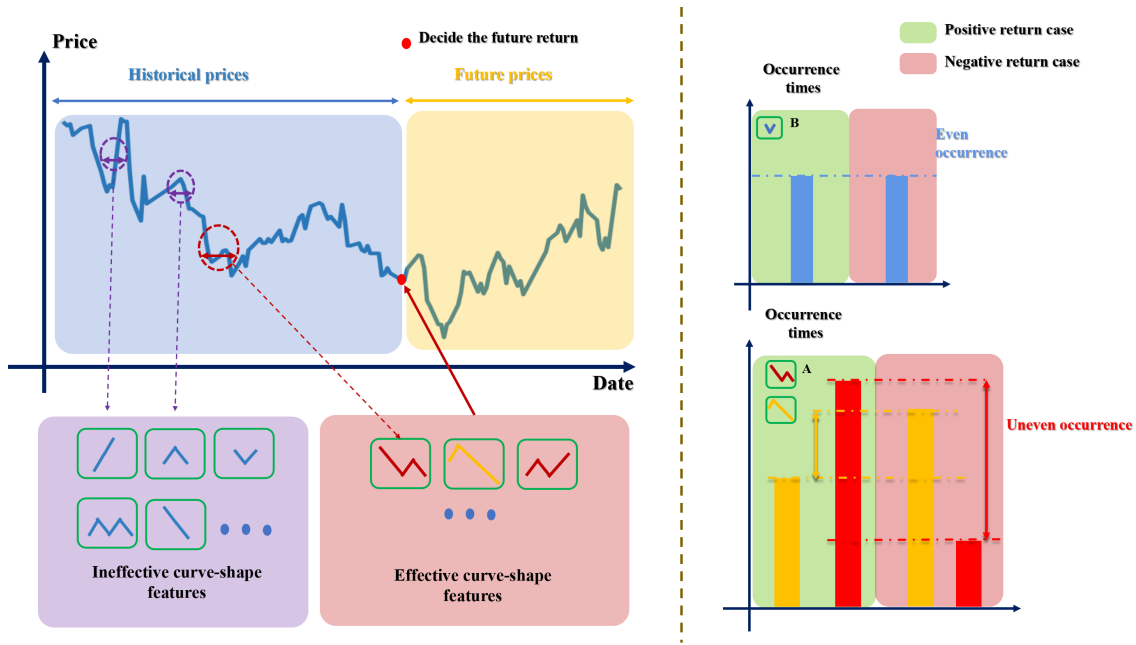}
\caption{Examples of effective curve-shape features.
Effective curve-shape features occur unevenly in the time series with 
positive and negative returns and decide the future returns.}
\label{efficient_feature1}
\end{figure}

\textit{The effective curve-shape features}
We can capture numerous different curve-shape features 
from historical financial data. For example, curve-shape features 
intercepted within different-size windows, as 
shown in Fig. \ref{efficient_feature1}.
The curve-shape features that are strongly correlated with 
the future return is the effective curve-shape features.
For example, the curve-shape feature A occurs at a higher frequency 
in a positive return history series, then, A is an effective curve-shape feature.
The curve-shape feature B occurs evenly in a positive and negative return history
series respectively, then, B is not an effective curve-shape feature.
A combination of effective curve-shape features can predict the future return.

\begin{figure}[H]
\centering
\includegraphics[width=1.0\textwidth]{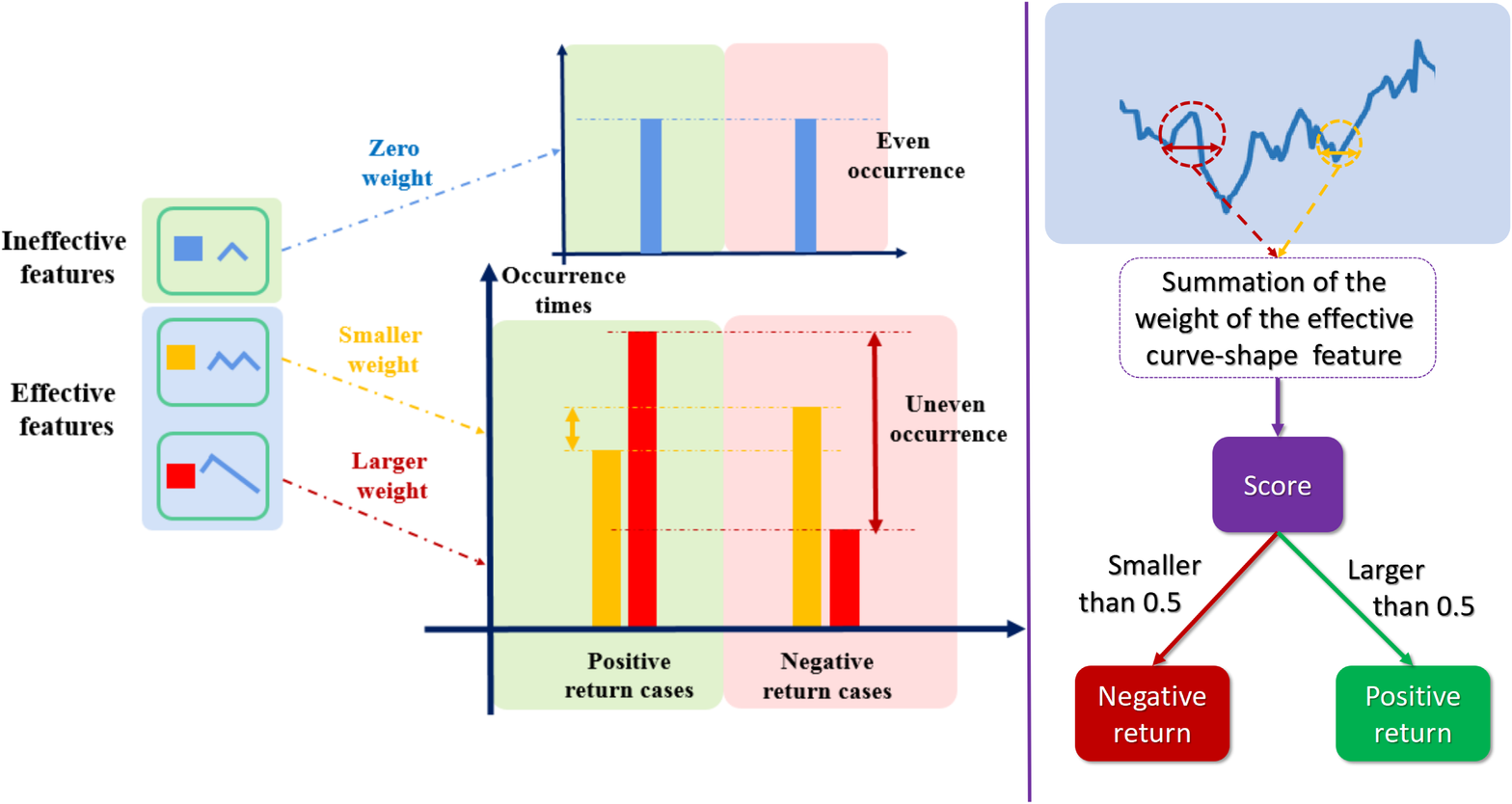}
\caption{Examples of effective curve-shape features with weights
and CSF mode series. The effective curve-shape features 
with higher degree of uneven occurrences have larger weights.
The summation of the weight of the effective curve-shape features 
in a time series, namely the score of the 
time series, decide the future return. For example, if the 
score beyond a threshold, the future return will be positive 
with a high probability.}
\label{efficient_feature2}
\end{figure}

\textit{The CSF mode series}. In the CSF mode series 
the future return can be extracted from the curve shapes of the historical
stock price.
We manually assign a weight on the curve-shape features. 
The summation of the weight 
of the features occur in a given historical stock prices decide
the future return. In the CSF mode series, those features 
with larger weight are effective curve-shape features, 
as show in Fig. \ref{efficient_feature2}.

\subsection{The NCSF mode series: the momentum-featured series}
\begin{figure}[H]
\centering
\includegraphics[width=1.0\textwidth]{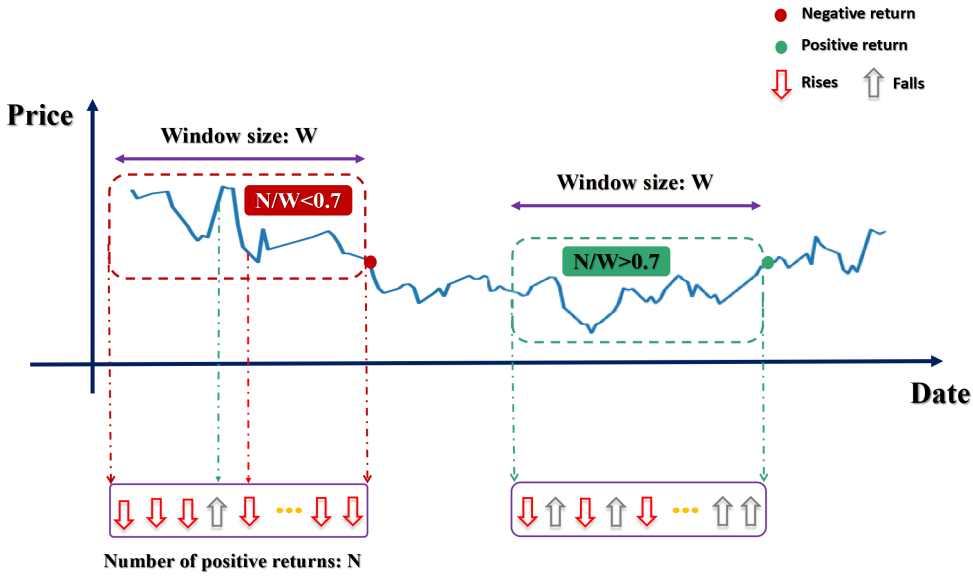}
\caption{An example of the momentum-featured series}
\label{momentum_feature}
\end{figure}

In this section, we manually design a NCSF mode series, where the future return
can not be extracted from the curve shape of the historical stock prices.
In this series, the future return is determined by the 
number of rises and falls in the historical data, For the sake of convenient,
we name it the momentum-featured series.
For example, for a historical data with in a fixed-size window, 
the ratio of the amount of rising data to the total amount of data is calculated. 
When the ratio exceeds a given value, say $0.7$, the future return will be positive 
with a high probability, as shown in Fig. \ref{momentum_feature}.
The momentum-featured series is just one of the NCSF mode series.

\subsection{The selected four kinds of series}
\begin{figure}[H]
\centering
\includegraphics[width=1.1\textwidth]{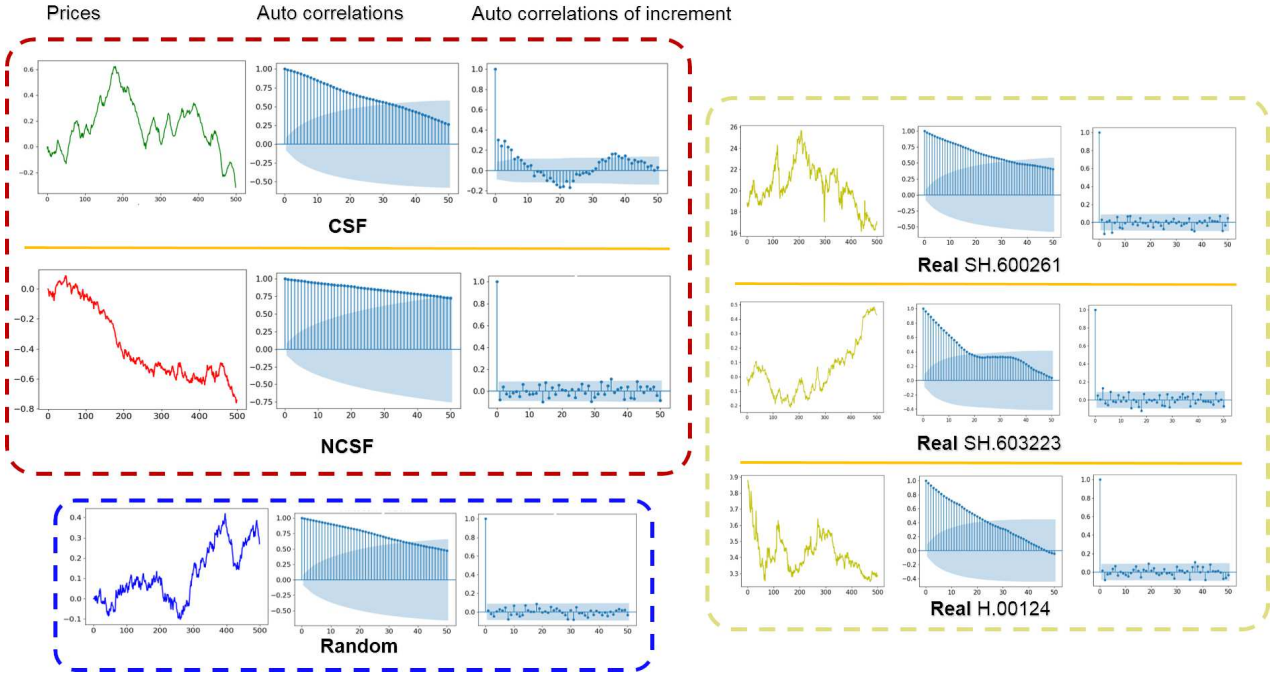}
\caption{A diagram of the prices and autocorrelation function of 
the four kinds of series including the CSF mode, NCSF mode, random generated, and real series. }
\label{four}
\end{figure}

In this section, we introduce the four kinds of series that are used in the
following experiment, and they are: (1) the CSF mode series,
(2) the momentum-featured series which is a typical NCSF mode series,
(3) the real stock series, and (4) the random generated series. As shown in Fig. \ref{four}.

In the series, the maximum size of the time window is $20$, that is, the next days return is 
decided by the prices in previous $20$ days.

\section{The methods and the experiment settings}
In this section, various of existing algorithms are applied on the
four kinds of series, including a new proposed statistical method, 
the existing machine learning methods,
and the existing deep learning methods.
The result shows that, the existing algorithms
are good at extracting the curve-shape features in the historical stock price. 
That is, the existing algorithms are effective for the CSF mode series
and are inappropriate for the NCSF mode series.

\subsection{The proposed statistical method for curve-shape features (the SM-CSF model) }
In this section, we propose a statistical method that is designed to extract the 
effective curve-shape features in the series. For the sake of convenient,
we name this method statistical method for curve-shape features (the SM-CSF model).
In this section, we give an introduction of the SM-CSF model.

In the SM-CSF model, we firstly collect all possible simplified curve-shape features in given window sizes, say
window sizes 4, 5, 6, and 7. Secondly, we count the occurrence times of each curve-shape features in 
the positive and negative samples. And the effective curve-shape features are selected according to 
their unevenly occurrence times. For example, if the feature A occurs $100$ times in the positive samples and $1000$
times in the negative samples, we consider A as an effective curve-shape feature.
Finally, the linear regression method is applied to find the weight that evaluates the
importance of the each effective curve-shape feature to the future return. 
That is, a weight is assigned to each effective features and the weight is evaluated by linear regression method \cite{islam2005multiple}.

\subsection{The existing algorithms: a brief review}
In this section, we give a brief review of the existing algorithms, including 
the statistical methods, machine learning methods,
and deep learning methods

\textit{Machine learning methods: support vector machine (SVM)} The SVM 
is a supervised learning model and related learning algorithm for 
data analysis in classification and regression analysis. 
Given a group of training instances, each training instance 
is marked as belonging to one or the other of the two categories. 
SVM training algorithm creates a model to assign the new instance 
to one of the two categories, making it a non probabilistic binary linear classifier \cite{osisanwo2017supervised}. 

\textit{Machine learning methods: random forest (RF)}
It can be regarded as a basic classification method of decision tree. 
The decision tree is composed of nodes and directed edges. The internal 
nodes represent feature attributes, and the external nodes (leaf nodes) 
represent categories.
RF is a bagging combination of different decision trees, which 
is based on decision tree.
The decision tree selects an optimal feature (maximum information gain ID3, 
maximum information gain ratio C4.5, minimum Gini index) from the feature 
set to branch, while the random forest selects the optimal feature from the 
randomly selected feature subset to branch.

\textit{Machine learning methods: multilayer perceptron (MLP)}
The MLP is a kind of forward structure artificial neural network, 
which maps a group of input vectors to a group of output vectors. 
MLP can be regarded as a directed graph, which is composed of multiple 
node layers, and each layer is fully connected to the next layer. 
Except for the input nodes, each node is a neuron with a nonlinear 
activation function. 

\textit{Machine learning methods: bayesian classifier}
Based on Naive Bayes formula, the maximum value of a posteriori probability is compared to classify. 
The calculation of a posteriori probability is obtained by the product of a priori probability and class conditional probability. 
A priori probability and class conditional probability are obtained by training data set.

\textit{Deep learning methods: the embedded convolution neural network (CNN)}
The convolutional neural network contains a feature extractor composed of convolution layer and subsampling layer. 
In convolution layer of convolution neural network, one neuron is only connected with some neighboring neurons. 
In a convolution layer of CNN, there are usually several feature maps. 
Each feature plane is composed of some neurons arranged in a rectangle. 
The neurons in the same feature plane share weights, and the weights shared here are convolution kernels. 
Convolution kernel is usually initialized in the form of random decimal matrix. 
In the process of network training, convolution kernel learn to get reasonable weights. 
Subsampling, also known as pooling, usually has two forms: mean pooling and Max pooling \cite{palaz2015analysis}. 
Convolution and subsampling greatly simplify the model complexity and reduce the model parameters.

We design an embedded convolution neural network \cite{cho2020combinatorial}. 
In this network, firstly, embedding the input series, 
sparsely representing the original data, 
converting each value in the series into vectors, learning into vector space. 
After that we convolute the embedded series by using 32 filters of size 5x7x9, 
then pooling them to extract the effective features. 
Finally, the output prediction value is calculated at the full connection layer.

\textit{Deep learning methods: the embedded long short-term memory neural network (LSTM)}
LSTM is an improved RNN. 
LSTM networks consist of LSTM units. LSTM unit is composed of cells having input, 
output and forget gates. These three gates regulate the information flow. With these 
features, each cell remembers the desired values over arbitrary time intervals. LSTM cells
combine to form layers of neural networks \cite{kim2019predicting}.

Based on the biLSTM, we structure an embedded biLSTM, 
where we embed the input series firstly, 
which converts each value in the series into a vector, 
this process can represent the series sparsely in vector space. 
Next,the embedded series is input from forward LSTM and backward LSTM in one time step. 
Forward LSTM and backward LSTM are calculated to get two sets of hidden vectors with valid features. 
Then two sets of hidden vectors are stitched together to get the final hidden state \cite{siami2019performance}. 
The model’s output layer calculates the predicted value based on parameters such as hidden state and weight, 
and the Loss function use the cross-entropy.

\subsection{The ground truth of the CSF and the NCSF mode series}
In this work, the CSF and NCSF series are generated manually.
According to the generation rules of the CSF and the NCSF series, we can easily
design a model that can extract the information of future return from such series. 
With this regard, we consider
the model as the model of ground truth, because it extracts the maximum degree of information 
from the CSF and the NCSF series. For the sake of convenient, we name them the ground truth model for the 
CSF series (GT-CSF)
and ground truth model for the NCSF model (GT-NCSF).

\textit{The GT-CSF model}. In the GT-CSF model, the pre-designed rule is used to design the model. For example, 
we know the weight for each curve-shape features and the threshold of the score. By directly calculating the 
score of the series and comparing it with the threshold, we can give the trend of future return.

\textit{The GT-NCSF model}. In the GT-NCSF model, the pre-designed rule is also used to design the model. For example, 
we know the ratio between ups and downs that decide the future return. By directly calculating the 
ratio of ups and downs and comparing it with the threshold, we can give the trend of future return.

\begin{figure}[H]
\centering
\includegraphics[width=1.0\textwidth]{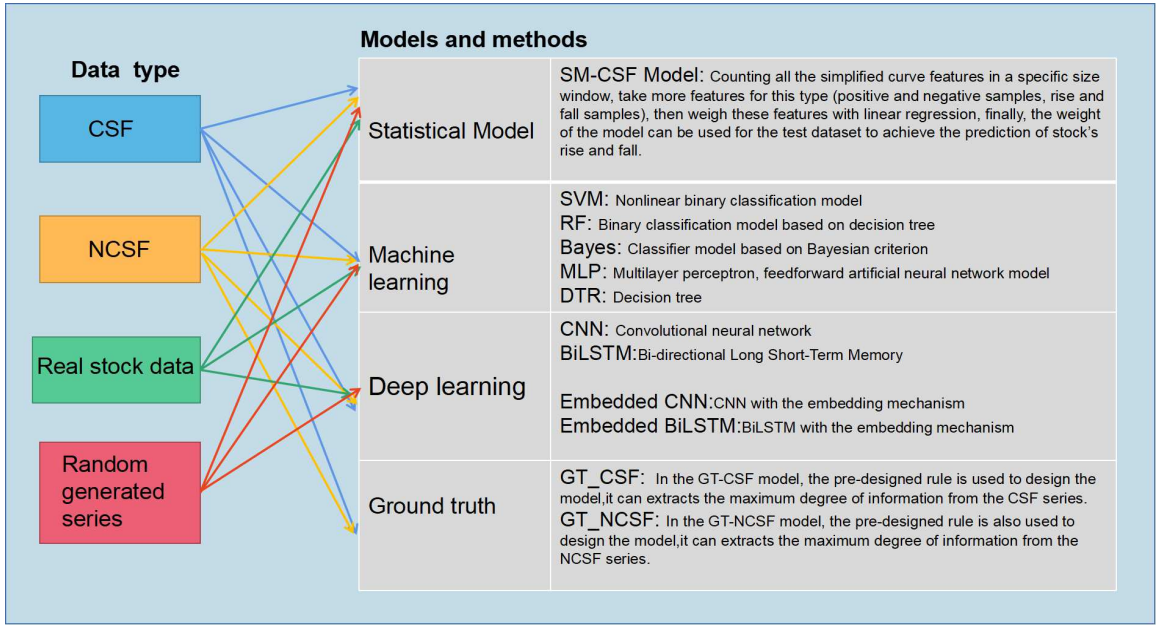}
\caption{An overview of the methods}
\label{methods}
\end{figure}

An overview of the methods is given in Fig. \ref{methods}.

\section{The results and analysis}
In this section, we show the results of different algorithms on the four series.

\subsection{The criteria} 
We give a criterion to judge the behavior of the algorithms as follows:
in a given test data set, we calculate the precision of the positive return cases in  
selected samples by each algorithm and compare it with that of the random 
selected samples.
The precision given by the ground truth gives the upper bound over all the algorithms.
For example, in the randomly selected samples, the precision
of the positive return cases is $0.52$. The ground truth is $0.75$ evaluates the 
maximum amount of the information of future return that is contained in the historical prices.
If the precision of selected samples by algorithm A
is obviously larger than $0.52$, then algorithm A is concluded to be effective. 
That is, algorithm A can extract the mode where the future return 
is correlated with the historical 
price. Otherwise, algorithm A is ineffective on the series.  

\subsection{The result of the CSF mode series} 
\begin{figure}[H]
\centering
\includegraphics[width=1.0\textwidth]{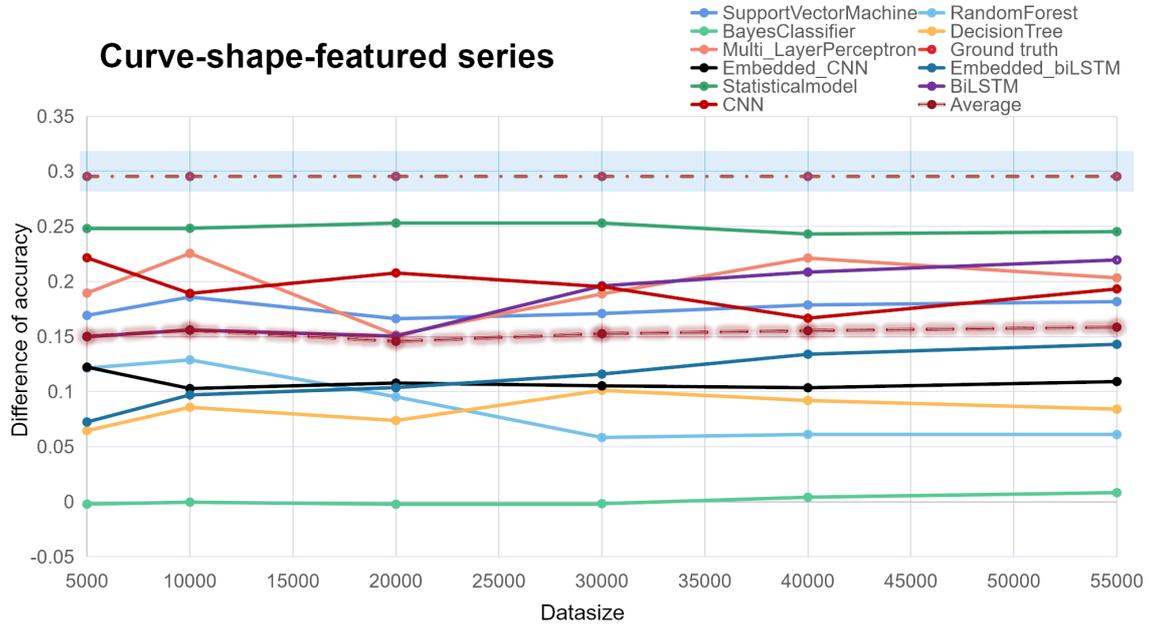}
\caption{The result of the CSF mode series}
\label{csf01}
\end{figure}
\subsection{The result of the NCSF mode series} 
\begin{figure}[H]
\centering
\includegraphics[width=1.0\textwidth]{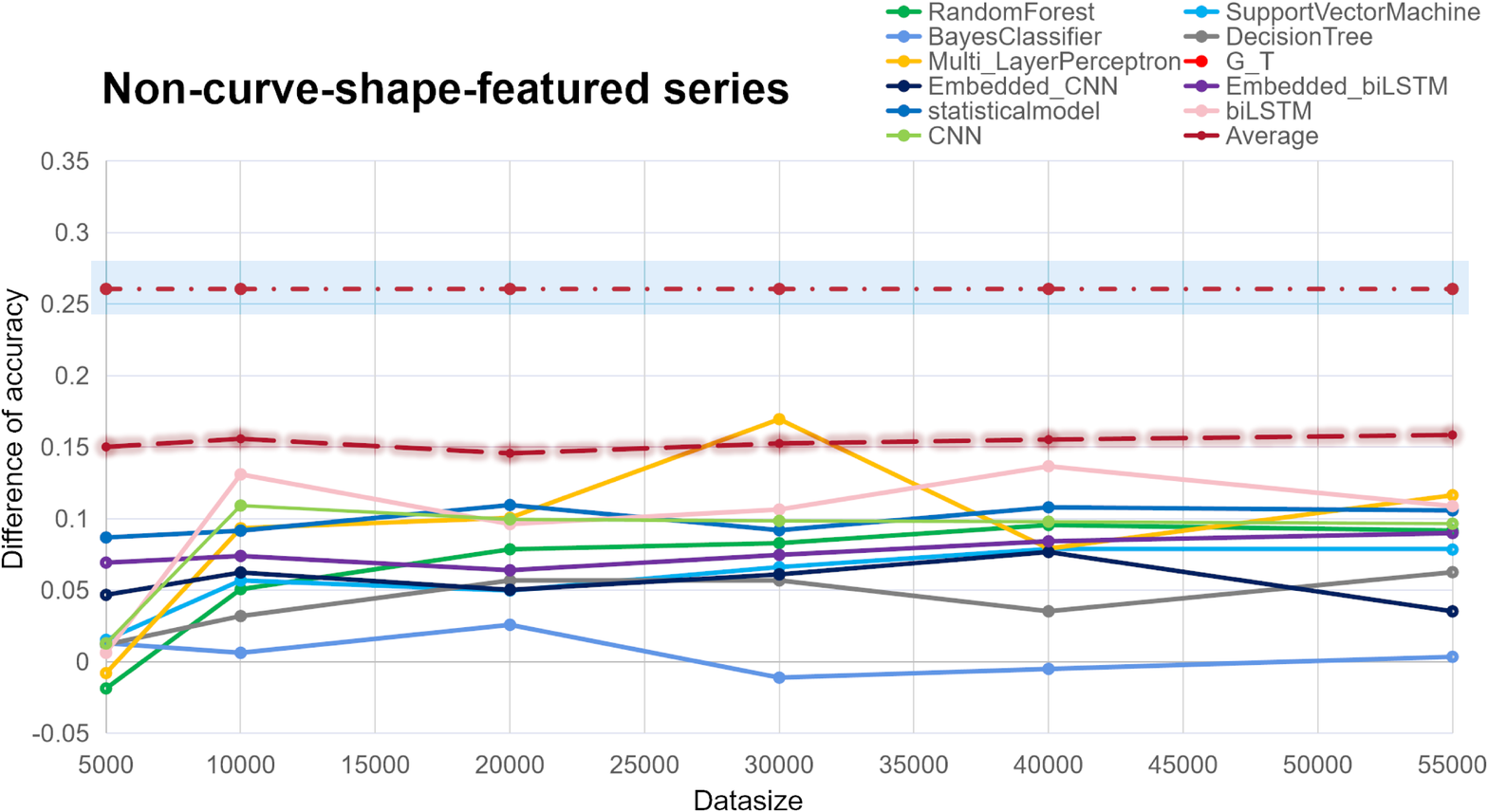}
\caption{The result of the NCSF mode series}
\label{ncsf01}
\end{figure}
\begin{figure}[H]
\centering
\includegraphics[width=1.0\textwidth]{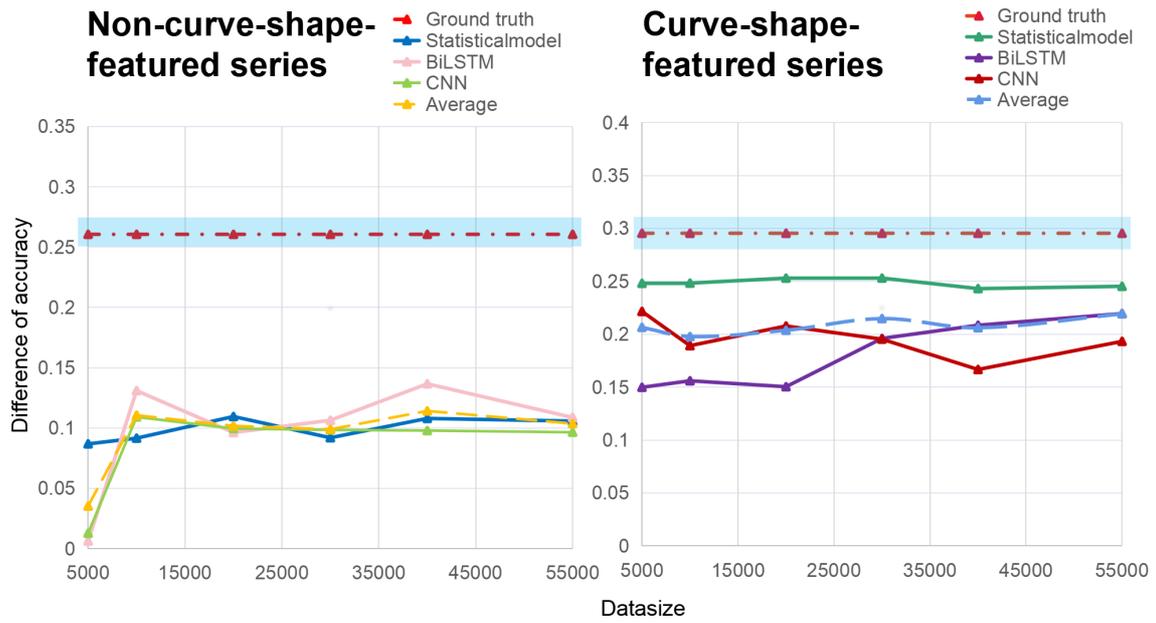}
\caption{The result of several models(deep learning models,statistical model)for CSF mode series}
\label{deep}
\end{figure}
From these different models, we put forward the first four with the best effect and put them in Figure 8. 
Through comparison, it can be observed that the 
deep learning model is more suitable for extracting the CSF, 
while the extraction effect for the NCSF is poor.
\subsection{The result of the real series} 
\begin{figure}[H]
\centering
\includegraphics[width=1.0\textwidth]{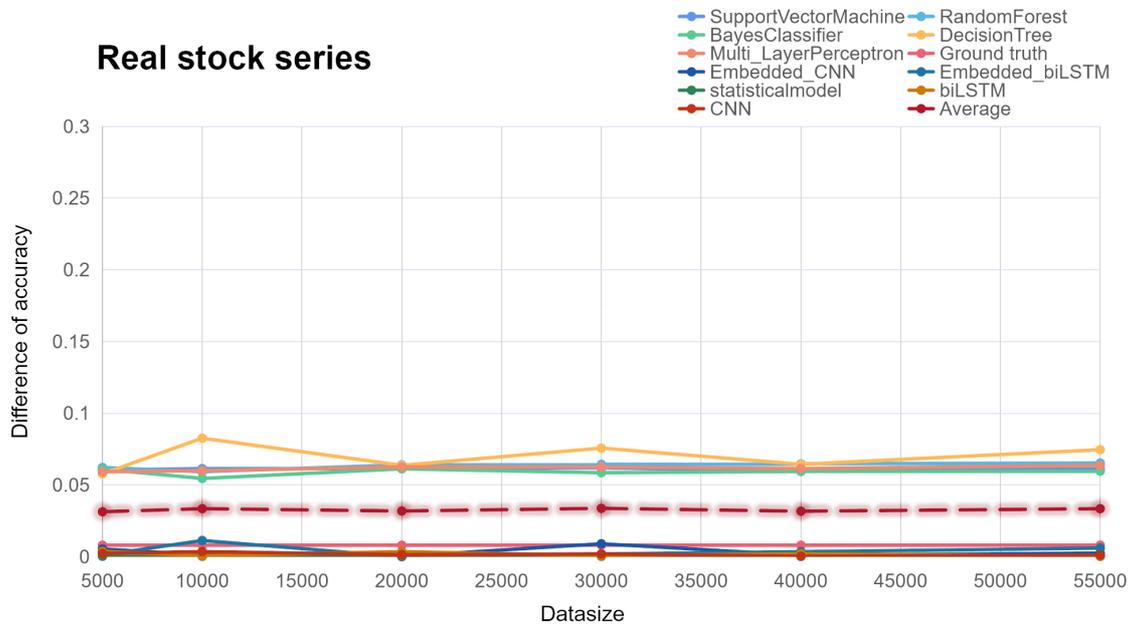}
\caption{The result of the real stock series}
\label{rss01}
\end{figure}

\subsection{The result of the $random_generated$ series} 
\begin{figure}[H]
\centering
\includegraphics[width=1.0\textwidth]{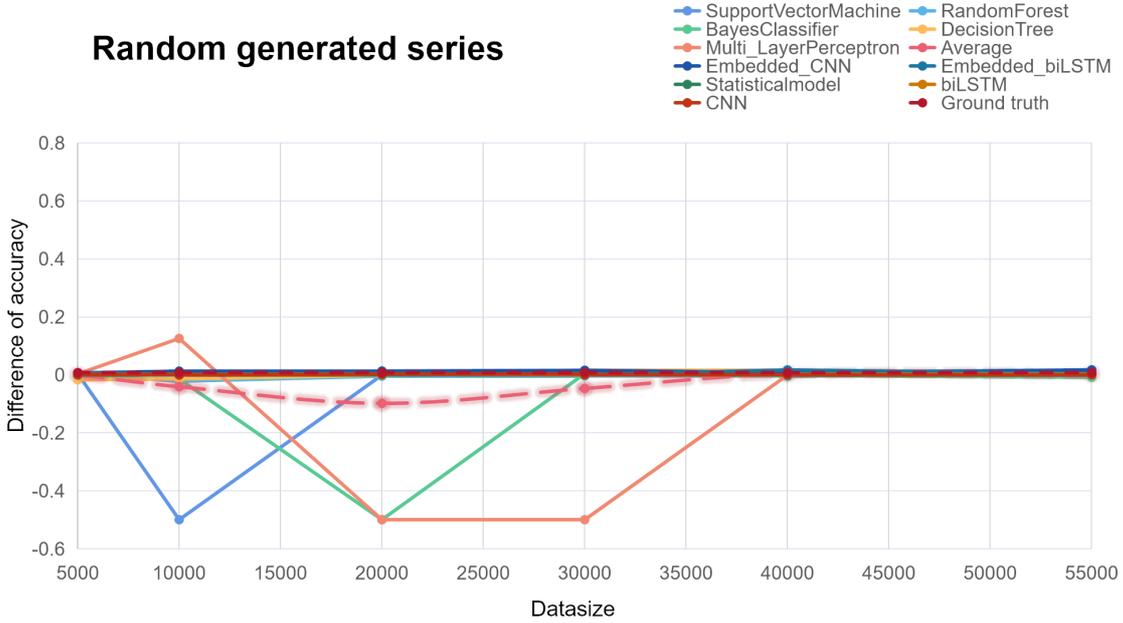}
\caption{The result of the $random_generated$ series}
\label{rgs01}
\end{figure}

\subsection{The analysis}
We conclude that in the real financial time series, less information 
of future return is contained in the curve-shape manner and more information 
is contained in the non-curve-shape manner. 
It points out that beyond the existing models, new models that can extract 
non-curve-shape features are needed in the financial time series analysis. 
We show that various kinds of existing machine learning models which are 
successful in the image recognition and natural language processing are 
inappropriate in financial time series analysis. 
We also point out the reason: various kinds of existing machine learning 
models are good at extracting the curve shape features and the information 
of the future return is not all contained in the curve shape features of historical data.

\section{Conclusions and outlook}

Through the above research results, we can see that statistical models, machine learning models, 
and deep learning models can give effective classification and prediction results for the curve-shape-featured data, 
however, for the non-curve-shape-featured data which does not contain the significant laws, 
these multi domain models cannot effectively classify them, that is also completely consistent with our speculation.

This is also sufficient to explain that in the financial data, due to a large number of combined 
curve-shape-features and non-curve-shape-features, the current forecasting methods cannot provide effective answers 
to the trend of future income. At the same time, we also pay attention to that in the financial market, 
different participants and researchers have their own subjective combination of characteristics, 
because according to the different time scales, a lot of combined curve-shape-features can be 
observed from the historical data, so everyone form their own judgment methods, such as those experienced investors. 
Therefore, the complex and diversified combination of curve-shape-features also affects the above 
model to judge the future income results differently.

Our breakthrough in this research is to prove the existence of effective curve-shape-features 
from complex financial data. 
These features of different curve-shape patterns contain effective information that affects future 
income results. According to the different combination methods and orders of various data-shape 
forms, the formed laws and speculative results also be different. In addition, the data are mixed 
with a large number of non-curve-shape-featured data (disorderly and random data). 
In the test of A-shares and Hong Kong stocks data, the test results of the three domain models 
are invalid, which indicates that the market is not effective under the premise of this theory, 
if we want to predict the future trend by analyzing the historical data, we can either analyze 
more historical data to extract enough effective features,
 so as to improve the more accurate basis for the future trend. Or we can only design more 
 complicated and more powerful models (more intelligent models), more accurate parameters, 
more complicated data processing and model optimization.

In this experiment, we use supervised models and discrimination models. 
These models can be understood as completely relying on data volume, just like use memory. 
Therefore, the data classification and prediction effect of curve-shape-features is better. 
For complex data that cannot identify the laws, they cannot make effective classification 
and prediction. So, we turn to the use of generative models in the next step: AE, GAN, etc.

\section{Acknowledgments}
We are very indebted to Prof. Wu-Sheng Dai, Guan-Wen Fang, and Yong-Xie for their encouragement. 
This work is supported by Yunnan Youth Basic Research Projects (202001AU070020) 
and Doctoral Programs of Dali University (KYBS201910).

%\appendix
%\section{Some title}
%Please always give a title also for appendices.

%\acknowledgments%崑仍
%%%%%%%%%%屎猟潤崩

%\begin{thebibliography}{99}

%\end{thebibliography}\endgroup

%\bibitem{a}
%Author, \emph{Title}, \emph{J. Abbrev.} {\bf vol} (year) pg.

%\bibitem{b}
%Author, \emph{Title},
%arxiv:1234.5678.

%\bibitem{c}
%Author, \emph{Title},
%Publisher (year).

% Please avoid comments such as "For a review'', "For some examples",
% "and references therein" or move them in the text. In general,
% please leave only references in the bibliography and move all
% accessory text in footnotes.

% Also, please have only one work for each \bibitem.

%\end{thebibliography}

%\bibliographystyle{JHEP}
%\bibliography{refs}% Produces the bibliography via BibTeX.

\end{document}